\documentclass[12pt]{article}
\usepackage{amsfonts}
\usepackage{amssymb}
\usepackage{graphics}
\begin{document}
{\LARGE \centerline{Two-particle Wigner functions} \centerline{in
a one-dimensional Calogero-Sutherland potential}}

\phantom{aaa}

\phantom{aaa}

{\large \centerline{A. Te\u{g}men; T. Altanhan and B. S. Kandemir}}%
\centerline{Physics Department, Ankara University, 06100 Ankara,
TURKEY}%
\centerline{{\it tegmen@science.ankara.edu.tr}} \centerline{{\it
altanhan@science.ankara.edu.tr}} \centerline{{\it
kandemir@science.ankara.edu.tr}}

\begin{abstract}
We calculate the Wigner distribution function for the
Calogero-Sutherland system which consists of harmonic and
inverse-square interactions. The Wigner distribution function is
separated out into two parts corresponding to the relative and
center-of-mass motions. A general expression for the relative
Wigner function is obtained in terms of the Laguerre polynomials
by introducing a new identity between Hermite and Laguerre
polynomials.

\end{abstract}
\noindent\rule{7in}{0.01in}
\section{Introduction}
\label{intro} Since the introduction of the Wigner function (WF)
in 1932 for inclusion of quantum corrections to classical
results\cite{Wigner32}, phase space representations of the quantum
mechanics have been a focus of continuing interest and found a
wide range of applications\cite{Lee95}. In this formalism one
defines a distribution function $W\left(q,p\right)$ of position
$q$ and momentum $p$ in such a way that to every normalized state
vector $\psi $ there corresponds a distribution function. In order
to define
the same physical system, $W$ should be a Hermitian form of $\psi$ , and if $%
W$ is integrated over $p$ it should give the proper probabilities
of the different values of $q$ or vice versa. Since the WF is a
probability distribution function one expects, as a natural
condition on $W(q,p)$, that it should be non-negative for all
values of $q$ and $p$: $W\left(q,p\right)\geq 0$. However
,Wigner\cite{Wigner79} proved that this result is incompatible
with the first two conditions and it is now a common practice to
work with a distribution function taking negative values for
certain $q$ and $p$ in the phase space.

The quantum Calogero-Sutherland model (CSM) having a quadratic
confining $q^{2}$ plus an inversely quadratic $1/q^{2}$
potentials\cite{Calogero71} has applications in a wide variety of
different areas of many body physics, due to the connection of its
variants and itself directly with the hierarchical fractional
quantum Hall effect\cite{Kawakami93}, free oscillators on a
circle\cite{Gurappa00}, the spectrum of the Chern-Simons matrix
model\cite{Karabali02}, short range Dyson model\cite{Ezung05}, and
Witten-Dijkgraff-Verlinde equation\cite{Belluci05}. Additionally,
many works have also been realized to construct its N-fermion
version\cite{Kilic93}, $W_{\infty}$ algebra
unification\cite{Hikami94}, shape invariance\cite{Efthimiou97},
generalized statistics\cite{Polychronakos99}, statistical
properties of quantum quasi-degenaracy\cite{Chalevabarti02},
equivalence to decoupled oscillators\cite{Gurappa99}.

In addition to allowing one to analyze the dynamics of quantum
systems entirely in phase space and thus to make comparison
between their classical and quantum evolutions, there is also
experimental interest on the measurements of WFs for certain
quantum systems to probe the predictions of quantum mechanics,
since WF contains complete quantum mechanical information as the
wave function or density matrix has. A number of experiments have
been reported where the measurement of WFs carried out for both
vacuum and quadrature-squeezed states of light\cite{Smithey93},
molecular vibrational states\cite{Dunn95}, various quantum states
of the motion of a harmonically trapped atom \cite{Leibfried96},
as well as for a massive particle wave packet\cite{Kurtsiefer97}.
Of particular importance are the experiments upon which the
negatives in WFs corresponding to Fock states\cite{Leibfried96}
and a superposition of macroscopically separated parts of matter
field\cite{Kurtsiefer97} have been observed. In this regard, due
to the fact that it is now possible to realize new quantum
mesoscopic devices such as quantum dots and quantum antidots by
various experimental techniques, the CSM  may serve as a model of
two non-interacting electrons with an individual quantum antidot
confined in a quantum wire or a stripe, where the repulsive
inverse-square quantum antidot potential acts as a scattering
center for electrons, or may be used as a one dimensional exactly
soluble band model or quantum dot arrays wherein the coupling
constant of the inverse-square potential of the CSM is chosen as
$0\geq g\geq-1/2$ in dimensionless units, i.e.,
attractive\cite{Sutherland71,Scarf58}. Moreover, very recently, Li
\textit{et al}.\cite{Li05} have solved CSM with pseudo-angular
momentum operator method, and they showed that, by discussing its
several variants, the radical equations of three dimensional
isotropic oscillator and hydrogen-like atom in both spherical and
parabolic coordinates, one dimensional three body problem and the
s-state of Morse potential all reduces to CSM. Therefore, WFs of
CSM may provide a solid basis for the discussions of transport
properties\cite{Ferry97} of the above mentioned nanostructures.
With these motivations, we study the WFs of CSM which has not only
a particular significance in itself, but also enables us to
understand the phase space picture of its variants and itself as
well.

The WF for two particles is defined by
\begin{eqnarray}
& &W\left( q_{1},q_{2};p_{1},p_{2}\right) =\frac{1}{\left( \pi
\hbar \right)
^{2}}\int_{-\infty }^{+\infty }\rm{d}y_{1}\int_{-\infty }^{+\infty }\rm{d%
}y_{2} \nonumber \\
& &\times \bar{\Psi} ^{\ast }\left( q_{1}+y_{1},q_{2}+y_{2}\right)
\bar{\Psi} \left( q_{1}-y_{1},q_{2}-y_{2}\right)\nonumber \\
& &\times \exp \left[ 2i\left( p_{1}y_{1}+p_{2}y_{2}\right) /\hbar
\right].  \label{1}
\end{eqnarray}
If we express the WF in terms of the center-of-mass and relative
coordinates through the relations
\begin{eqnarray}
q_{1}+q_{2} & = & 2Q \;,\qquad q_{1}-q_{2} = q \nonumber \\
p_{1}+p_{2} & = & 2P \;,\qquad p_{1}-p_{2} = p \nonumber \\
y_{1}+y_{2} & = & 2Y  \;,\qquad y_{1}-y_{2} = y \nonumber
\end{eqnarray}
then, provided that the interparticle potential depends on the
relative coordinates, equation ~(\ref{1}) becomes
\begin{eqnarray}
& & W\left( q,Q;p,P\right) =\frac{1}{\left( \pi \hbar \right)
^{2}}\int_{-\infty }^{+\infty }\rm{d}y\int_{-\infty }^{+\infty
}\rm{d}Y \nonumber \\
& &\times \Psi ^{\ast }\left( q+y\right) \Psi^{\ast }\left(
Q+Y\right) \Psi \left( q-y\right) \Psi (Q-Y)\nonumber \\
& &\times \exp \left[ i\left( 4PY+py\right) /\hbar \right],
\label{2}
\end{eqnarray}
since the solution of the corresponding Schr\"{o}dinger equation
can be represented by a product of two functions, one for the
center-of-mass and other for the relative coordinates. Then,
Eq.~(\ref{2}) can be separated
as  and relative WFs, $W(q,p)$ and $%
W\left(Q,P\right) $, respectively. We can therefore define the WFs
for the center-of-mass and relative motions in the form
\begin{eqnarray}
W\left( Q,P\right) &=&\frac{1}{\pi \hbar }\int_{-\infty }^{+\infty }\rm{d}%
Y\ \Psi ^{\ast }\left( Q+Y\right) \ \Psi \left( Q-Y\right)\exp
\left[
4iPY/\hbar \right] ,  \label{3} \\
W\left( q,p\right) &=&\frac{1}{\pi \hbar }\int_{-\infty }^{+\infty }\rm{d}%
y\ \Psi ^{\ast }\left( q+y\right) \ \Psi \left( q-y\right) \exp
\left[ ipy/\hbar \right] ,  \label{4}
\end{eqnarray}
respectively.  It should be noted that, while $\bar{\Psi}$'s in
Eq.~(\ref{1}) represent the two body wavefunctions, $\Psi$'s in
Eq.~(\ref{2}-4) are single particle wave functions. Now, it is
possible to present a coupled system of linear partial
differential equations corresponding to the above defined WF,
which requires the direct computation without solving the wave
functions\cite{Hug98}. The Wigner representation is very
convenient for studying quantum systems with Hamiltonians that
include quadratic coordinates and momenta, since in this case the
Wigner distribution function represents a good approximate
description of the dynamics involved. The method, however, is not
easy to handle when the potential contains higher order powers of
coordinates, since this case comprises a differential equation for
the Wigner function with terms as much as the number of the order
. In recent years there has been a number of works to calculate
the WF for various type of potentials: Infinite
square well\cite{Belloni04}, a double well potential\cite{Novaes03}, the P%
\"{o}sch-Teller potential\cite{Bund00}, the Morse
oscillator\cite{Lee82}, a
quantum damped oscillator\cite{Akhundova95}, the hydrogen atom\cite{Nouri98}%
, the rotational motion of a spherical top\cite{Nasyrov99} are
notable applications. A discrete WF for non-relativistic quantum
systems with one degree of freedom has been developed in finite
dimensional phase space and applied to a few simple
system\cite{Hakioglu00}.

The layout of this paper is  as follows: In Sec.~\ref{sec:2}, we
discuss WFs for the center-of-mass and relative motions, and
obtain general expressions in terms of Laguerre polynomials. In
Sec.~\ref{sec:3}, we derive a new identity between Hermite and
Laguerre polynomials to obtain a compact form for the WF of the
relative part, and plot some of them for a few states to give an
idea on  their phase space behaviors.
\section{Theory}\label{sec:2}
The Hamiltonian describing two particles interacting pairwise by
the Calogero-Sutherland potential is given by
\begin{equation}
H=\sum\limits_{i=1}^{2}\left[ -\frac{\hbar ^{2}}{2m}\frac{\partial ^{2}}{%
\partial q_{i}^{2}}+\frac{1}{2}m\omega _{\bullet }^{2}q_{i}^{2}+\frac{1}{2}%
\sum\limits_{j\neq i}^{2}U\left( \left\vert q_{i}-q_{j}\right\vert \right) %
\right],  \label{5}
\end{equation}
where the second term is the confining potential and
\begin{eqnarray}
U\left( \left\vert q_{i}-q_{j}\right\vert \right) = \left[ m\omega
_{0}^{2}\left( q_{i}-q_{j}\right) ^{2}+2g/\left(
q_{i}-q_{j}\right) ^{2}\right] /2 \nonumber
\end{eqnarray}
simulates further interactions between two particles. If we now
use the
above defined center-of-mass and relative coordinates, then the relevant Schr%
\"{o}dinger equation is separated out as a center-of-mass
equation, which is a 1D harmonic oscillator equation
\begin{equation}
\left[ -\frac{\hbar
^{2}}{2M}\frac{d^{2}}{dQ^{2}}+\frac{1}{2}M\omega _{\bullet
}^{2}Q^{2}\right] \Psi \left( Q\right) =E_{Q}\Psi \left( Q\right)
, \label{6}
\end{equation}
and the Calogero-Sutherland system
\begin{equation}
\left[ -\frac{\hbar ^{2}}{2\mu
}\frac{d^{2}}{dq^{2}}+\frac{1}{2}\mu \omega
^{2}q^{2}+\frac{g}{q^{2}}\right] \Psi \left( q\right) =E_{q}\Psi
\left( q\right) ,  \label{7}
\end{equation}%
where we have defined
$\omega^{2}=\omega_{\bullet}^{2}+2\omega_{0}^{2}$ as a hybrid
frequency. In Eqs.~(\ref{6})-(\ref{7}), $M$ and $\mu$ are total
and reduced masses, and are given by $2m$ and $m/2$, respectively.

The Wigner function corresponding to the center-of-mass motion
defined by Eq.~(\ref{3}) through the solution of Eq.~(\ref{6}) is
well-known, and given by\cite{Lee95}
\begin{eqnarray}
W_{\ell }\left( Q,\widetilde{P}\right)&=&\frac{\left( -1\right)
^{\ell }}{\pi \hbar }\exp \left[ -M\omega_{\bullet } Q^{2}/\hbar
-\widetilde{P}^{2}/M\omega_{\bullet } \hbar %
\right] \nonumber \\
& &\times L_{\ell }\left( \frac{2M\omega_{\bullet } }{\hbar }Q^{2}+\frac{2\widetilde{P}%
^{2}}{\hbar M\omega_{\bullet } }\right) ,  \label{8}
\end{eqnarray}%
where $\widetilde{P}=2P$ is used for the sake of comparison with
the results presented in the associated literature, and $\ell$
takes values $0,1,2,\ldots$. Although it is possible to obtain the
WFs for the simple harmonic oscillator (Eq.~(\ref{8})) in various
ways, for example, by using algebraic methods or by solving
ordinary differential equations of WF \cite{Curtright}, the WF
corresponding to the relative motion resulting in the CS system
cannot be obtained by either methods. Therefore, we are compelled
to obtain the corresponding Wigner
function through solving Eq.~(%
\ref{7}). First, we make a change of variable by $z=(\mu \omega
/\hbar )^{1/2}q$ , which transforms Eq.~(\ref{7}) into
\begin{equation}
\Psi ^{\prime \prime }+\left( 4n+2\beta +2-z^{2}+\frac{1/4-\beta ^{2}}{z^{2}}%
\right) \Psi =0,  \label{9}
\end{equation}%
where the new parameters are given by
\begin{eqnarray}
E_{q}=\hbar \omega \left( 2n+\beta +1\right)\;,\;1/4-\beta
^{2}=-2\mu g/\hbar ^{2}\nonumber
\end{eqnarray}
with $n=0,1,2,\ldots$ and $g\geq-\hbar^{2}/8\mu$. It should be
noted that these are the energy levels of one-dimensional
isotropic harmonic oscillator with odd quantum numbers shifted by
an amount $\left( \beta -1/2\right)
\hbar \omega $. The solution to Eq.~(\ref{9}) can then be written\cite%
{Perelomov86} in terms of the Laguerre polynomials
\begin{equation}
\Psi _{n}\left( q\right) =C_{n}\ b^{\alpha /2}q^{\alpha }\exp
\left[ -bq^{2}/2\right] \ L_{n}^{\alpha -1/2}\left( bq^{2}\right)
,  \label{10}
\end{equation}%
where $\alpha=\beta+1/2$, $b=\mu \omega /\hbar $ and the normalization constant is given by $%
C_{n}=b^{1/4}\left[ n!/\Gamma \left( n+\alpha +1/2\right) \right]
^{1/2}$. We can now build the associated WFs for the relative
motion (RM) with the wave functions given by Eq.~(\ref{10})
according to the definition of Eq.~(\ref{4}), which results in
\begin{eqnarray}
W_{n\alpha }\left( q,\widetilde{p}\right)&=&\frac{\left\vert
C_{n}\right\vert ^{2}}{\pi \hbar }\ b^{\alpha }\exp \left[
-bq^{2}\right] \int_{-\infty }^{+\infty }\rm{d}y\ \left(
q^{2}-y^{2}\right) ^{\alpha }\nonumber \\
& &\times \exp \left[
-by^{2}\right] \ L_{n}^{\alpha -\frac{1}{2}}\left[ b\left( q+y\right) ^{2}%
\right] \nonumber \\
& &\times L_{n}^{\alpha -\frac{1}{2}}\left[ b\left( q-y\right)
^{2}\right] \ \exp \left( 2i\widetilde{p}y/\hbar \right) ,
\label{11}
\end{eqnarray}%
where $p=2\ \widetilde{p}$ is used. If we use the binomial
expansion of
\begin{eqnarray}
\left( a+d\right) ^{\alpha }=\sum\limits_{\beta =0}^{\alpha }
\left(
\begin{array}{c}
\alpha \\ \beta
\end{array}
\right) a^{\alpha -\beta }\ d^{\beta },\nonumber
\end{eqnarray}
and series expansion of the Laguerre polynomials
\begin{eqnarray}
L_{n}^{k}\left( x\right) =\sum\limits_{m=0}^{n}\frac{\left( -1\right) ^{m}}{%
m!} \left( \begin{array}{c} k+n \\ n-m
\end{array}\right)\,x^m \nonumber
\end{eqnarray}%
then Eq.~(\ref{11}) is expressed in the form
\begin{eqnarray}
&& W_{n\alpha }\left( q,\widetilde{p}\right) = \frac{\left\vert
C_{n}\right\vert ^{2}}{\pi \hbar }\ b^{\alpha }\exp \left[
-bq^{2}\right] \nonumber \\
&& \times \sum\limits_{m}^{n}\sum\limits_{r}^{n}\frac{\left(
-1\right) ^{m+r}}{m!r!} \left(
\begin{array}{c} \alpha-\frac{1}{2}+n \\ n-m
\end{array}\right)
\left( \begin{array}{c} \alpha-\frac{1}{2}+n \\ n-r
\end{array}\right)
b^{m+r}\nonumber \\
&& \times \sum\limits_{\beta =0}^{\alpha }\left( -1\right) ^{\beta
} \left( \begin{array}{c}\alpha \beta
\end{array}\right) q^{2\alpha -2\beta }\sum\limits_{\mu =0}^{2m}
\left(
\begin{array}{c}2m  \\ \mu \end{array}\right)
q^{2m-\mu }\nonumber \\
&& \times \sum\limits_{\rho =o}^{2r}\left( -1\right) ^{\rho }
\left(
\begin{array}{c} 2r \\ \rho  \end{array}\right) q^{2r-\rho }\
F_{2\beta +\mu +\rho }\left( \widetilde{p}\right) , \label{12}
\end{eqnarray}%
where $F_{2\beta +\mu +\rho }\left( \widetilde{p}\right) $ is
given by the following integral
\begin{equation}
F_{2\beta +\mu +\rho }\left( \widetilde{p}\right) =\int_{-\infty
}^{+\infty }dy\ y^{2\beta +\mu +\rho }\exp \left[
-by^{2}+2i\widetilde{p}y/\hbar \right] .  \label{13}
\end{equation}%
It is easy to show that this last integral can be expressed in
terms of the Hermite polynomial as follows:
\begin{equation}
F_{n}\left( \widetilde{p}\right) =b^{-\left( n+1\right) /2}\frac{\sqrt{\pi }%
}{2^{n}\left( -i\right) ^{n}}\exp \left[ -\widetilde{p}^{2}/b\hbar ^{2}%
\right] \ H_{n}\left( \frac{\widetilde{p}}{\hbar \sqrt{b}}\right)
, \label{14}
\end{equation}%
with $n=2\beta +\mu +\rho $\cite{Gradshteyn00}. Furthermore, the
use of the relation\cite{Gradshteyn00} $H_{n}\left( u\right)
=\left( -1\right) ^{n}\ e^{u^{2}}\partial _{u}^{n}\ e^{-u^{2}}$
for the Hermite polynomials reduces the above expression to
\begin{equation}
F_{n}\left( \widetilde{p}\right) =\sqrt{\frac{\pi }{b}}\left(
-\frac{i\hbar }{2}\ \partial _{\widetilde{p}}\right) ^{n}\
e^{-\widetilde{p}^{2}/b\hbar ^{2}},  \label{15}
\end{equation}%
where $\partial _{\widetilde{p}}$ represents the differentiation
with respect to $\widetilde{p}$. Hence, the WFs for the RM
becomes, with these new definitions,
\begin{eqnarray}
&& W_{n\alpha }\left( q,\widetilde{p}\right)=\frac{\left\vert
C_{n}\right\vert ^{2}}{\sqrt{\pi b}\hbar }\ \exp \left[
-bq^{2}\right]\nonumber \\
&& \times \sum\limits_{m=0}^{n}\frac{\left( -1\right)
^{m}}{m!}%
\left( \begin{array}{c}\alpha -\frac{1}{2}+n \\n-m
\end{array}\right)
b^{m}\ \left( q-\frac{i\hbar }{2}\ \partial _{\widetilde{p}%
}\right) ^{2m} \nonumber \\
&& \times \sum\limits_{r=0}^{n}\frac{\left( -1\right) ^{n}}{n!}
\left( \begin{array}{c}\alpha-\frac{1}{2}+n \\n-r
\end{array}\right)
b^{r}\ \left( q+\frac{i\hbar }{2}\ \partial _{\widetilde{p%
}}\right) ^{2r}\nonumber \\
&& \times b^{\alpha }\left( q^{2}+\frac{\hbar ^{2}}{4}\partial _{%
\widetilde{p}}^{2}\right) ^{\alpha }e^{-\widetilde{p}^{2}/b\hbar
^{2}}. \label{16}
\end{eqnarray}
By using once again series expansion of the Laguerre polynomials,
it is possible to express Eq.~(\ref{16}) in an implicit form as
well
\begin{eqnarray}
&& W_{n\alpha }\left( q,p\right) =\frac{n!}{\sqrt{\pi }\hbar
\Gamma \left( n+\alpha +1/2\right) }e^{-\mu \omega
q^{2}/\hbar}\nonumber \\
&& \times L_{n}^{\alpha -1/2}\left[ \frac{\mu \omega }{\hbar
}\left( q-i\hbar \ \partial _{p}\right) ^{2}\right] L_{n}^{\alpha
-1/2}\left[ \frac{\mu \omega }{\hbar }\left( q+i\hbar \ \partial
_{p}\right) ^{2}\right]\nonumber \\%
&&\times \left[ \frac{\mu \omega }{\hbar }\left( q^{2}+\hbar
^{2}\partial _{p}^{2}\right) \right] ^{\alpha }\ e^{-p^{2}/4\mu
\omega \hbar }. \label{17}
\end{eqnarray}%
If we now define a unit of dimension by $l=\sqrt{\hbar /m\omega
_{\bullet }}$ with $\mu =m/2$ and $M=2m$, then we can make
positions and momenta
dimensionless by $\overline{Q}=Q/l$ , $\overline{P}=lP/\hbar $, $\overline{q}%
=q/l$ , $\overline{p}=lp/\hbar $ \ , and frequency by $\overline{\omega }%
=\omega /\omega _{\bullet }$. Hence, the relevant WFs for the
center-of-mass and relative motions, Eqs.~(\ref{8}) and
(\ref{17}), become
\begin{equation}
\widetilde{W}_{\ell }\left( \overline{Q},\overline{P}\right)
=\left( -1\right) ^{\ell }\exp \left[
-2\overline{Q}^{2}-2\overline{P}^{2}\right] \ L_{\ell }\left(
4\overline{Q}^{2}+4\overline{P}^{2}\right) ,  \label{18}
\end{equation}
and
\begin{eqnarray}
&& \widetilde{W}_{n\alpha }\left( \overline{q},\overline{p}\right) =\frac{\sqrt{%
\pi }n!}{\Gamma \left( n+\alpha +1/2\right) }e^{-\overline{\omega }\overline{%
q}^{2}/2}\nonumber \\
&& \times L_{n}^{\alpha -1/2}\left[ \frac{\overline{\omega
}}{2}\left( \overline{q}-i\ \partial \overline{_{p}}\right)
^{2}\right] L_{n}^{\alpha
-1/2}\left[ \frac{\overline{\omega }}{2}\left( \overline{q}+i\ \partial _{%
\overline{p}}\right) ^{2}\right]\nonumber \\
&& \times \left[ \frac{\overline{\omega }}{2}\left(
\overline{q}^{2}+\partial _{\overline{p}}^{2}\right) \right] ^{\alpha }\ e^{-%
\overline{p}^{2}/2\overline{\omega }},  \label{19}
\end{eqnarray}
respectively, where we have denoted $\pi \hbar W_{\ell }\left(
\overline{Q},\overline{P} \right) $ and $\pi \hbar W_{n\alpha
}\left( \overline{q},\overline{p}\right)$ as $\widetilde{W}_{\ell
}\left( \overline{Q},\overline{P}\right) $ and $
\widetilde{W}_{n\alpha }\left( \overline{q},\overline{p}\right) $,
respectively.
\section{Results and Discussion}
\label{sec:3}

FIGs. 1 and 2 show, respectively, the WFs $%
\widetilde{W}_{02}$ and $\widetilde{W}_{03}$ for the relative
motion given by Eq.~(\ref{19}) as functions of dimensionless
position $\overline{q}=q/l$ and momentum $\overline{p}=lp/\hbar $
for two different dimensionless frequency values,
$\overline{\omega }=\omega /\omega _{\bullet }=1$ and $3$. Contour
plots showing the projections of the relevant WF onto
$(\overline{q},\overline{p})$ plane are also shown in these
figures. In other words, each contour is a slice of given WF in
the $(\overline{q},\overline{p})$ plane. It should be noted that,
while $\overline{\omega }=1$ corresponds to the case
$\overline{\omega}_{0}=0$, $\overline{\omega}=3$ corresponds to
switch on $\overline{\omega}_{0}$ to the value
$\overline{\omega}_{0}=1$, which causes localization in
$\overline{q}$\cite{Palacios04,Kandemir}. In addition to this
pattern, delocalization in $\overline{p}$ is observed in both
figures. In other words, in FIG. 1(b) and FIG. 2(b), there the
dips and peaks of WFs in $\overline{q}$ are shifted towards the
smaller $\overline{q}$ values compared with those in FIG. 1(a) and
FIG. 2(a), whereas those of WFs in $\overline{p}$ are shifted
towards higher $\overline{p}$ values.

Having obtained a general expression for the WF of two interacting
particles we now distinguish between the cases $g=0$ and $g\neq
0$, and deal with each case separately. This allows us to verify
the consistency of the WFs obtained above with those found in the
literature. In order to see this, we
need to set $g=0$ first. In  this case, we have only the solutions with $%
\beta =+1/2$ and $-1/2$ corresponding to $\alpha =1$ and $0$,
respectively. We obtain, for $n=0,1,2,3,\ldots $,
\begin{eqnarray}
\widetilde{W}_{n1}\left( \overline{q},\overline{p}\right) &=&-\frac{\sqrt{\pi }%
\left( 2n+1\right) !}{n!\ 2^{2n+1}\ \Gamma \left( n+3/2\right) }\exp \left( -%
\frac{\overline{\omega}\overline{q}^{2}}{2}-\frac{\overline{p}^{2}}{2%
\overline{\omega }}\right)
\ L_{2n+1}\left( \overline{\omega }\overline{q}%
^{2}+\frac{\overline{p}^{2}}{\overline{\omega } }\right)
\label{20}
\end{eqnarray}%
which are the relative WFs corresponding to eigenvalues of the
harmonic oscillators with $2n+1$ eigenvalues. To obtain
Eq.~(\ref{20}), we have used the fact that every power of
$\overline{q}^{2}+\partial _{\overline{p}}^{2}$ commutes with
Laguerre polynomials with argument of $\overline{q}\mp i\
\partial _{\overline{p}}$ in Eq.~(\ref{19}), and the identity%
\begin{eqnarray}
\sqrt{\frac{\overline{\omega }}{2}}\left( \overline{q}\mp i\partial _{%
\overline{p}}\right) L_{n}^{1/2}\left[ \frac{\overline{\omega
}}{2}\left( \overline{q}\mp i\ \partial _{\overline{p}}\right)
^{2}\right]
=\frac{%
(-1)^{n}}{2^{2n+1}n!}H_{2n+1}\left[ \sqrt{\frac{\overline{\omega }}{2}}%
\left( \overline{q}\mp i\ \partial _{\overline{p}}\right) \right],
\label{21}
\end{eqnarray}%
and we have derived a new identity  between Hermite and Laguerre
polynomials in the form of
\begin{eqnarray}
H_{n}(\overline{u}+\frac{i}{2}\partial _{\overline{\upsilon }})H_{n}(%
\overline{u}-\frac{i}{2}\partial _{\overline{\upsilon }})e^{-\overline{%
\upsilon }^{2}}
=(-1)^{n}2^{n}n!L_{n}\left[ 2\left( \overline{u}^{2}+%
\overline{\upsilon }^{2}\right) \right] e^{-\overline{\upsilon
}^{2}}. \label{22}
\end{eqnarray}%
The proof of Eq.~(\ref{22}) can easily be done by using standard
relations among these polynomials.
In case of $\alpha =0$, we proceed as before by formally using Eqs.~(\ref{21}%
) and ~(\ref{22})  to give
\begin{eqnarray}
\widetilde{W}_{n0}\left( \overline{q},\overline{p}\right) &=&\frac{\sqrt{\pi }%
\left( 2n\right) !}{n!\ 2^{2n}\ \Gamma \left( n+1/2\right) }\exp \left( -%
\frac{\overline{\omega }\overline{q}^{2}}{2}-\frac{\overline{p}^{2}}{2%
\overline{\omega }}\right) L_{2n}\left( \overline{\omega }\overline{q}^{2}+%
\frac{\overline{p}^{2}}{\overline{\omega }}\right)  \label{23}
\end{eqnarray}%
which are the relative WFs corresponding to eigenvalues of the
harmonic oscillators with $2n$ eigenvalues, again with
$n=0,1,2,3,\ldots $. A more general expression for these two cases
can be found by noticing that arrangement of the coefficients in
Eqs.~(\ref{20}) and ~(\ref{23}) leads to the
pair of equations%
\begin{eqnarray}
\widetilde{W}_{n}\left( \overline{q},\overline{p}\right) =\left\{
\begin{array}{c}
\ -\exp \left( -\frac{\overline{\omega }\overline{q}^{2}}{2}-\frac{\overline{%
p}^{2}}{2\overline{\omega }}\right) \ L_{2n+1}\left( \overline{\omega }%
\overline{q}^{2}+\frac{\overline{p}^{2}}{\overline{\omega } }\right), \\
+\exp \left( -\frac{\overline{\omega }\overline{q}^{2}}{2}-\frac{\overline{p}%
^{2}}{2\overline{\omega }}\right) \ L_{2n}\left( \overline{\omega }\overline{%
q}^{2}+\frac{\overline{p}^{2}}{\overline{\omega }}\right),%
\end{array}%
\right.\nonumber
\end{eqnarray}%
or they may be combined into the form of
\begin{equation}
\widetilde{W}_{n}\left( \overline{q},\overline{p}\right)
=(-1)^{n}\exp
\left( -\frac{\overline{\omega }\overline{q}^{2}}{2}-\frac{\overline{p}^{2}}{%
2\overline{\omega }}\right) \ L_{n}\left( \overline{\omega }\overline{q}^{2}+%
\frac{\overline{p}^{2}}{\overline{\omega }}\right).\label{24}
\end{equation}%
The WFs constructed from the products of Eq.~(\ref{24})  with the
Eq.~(\ref{18}) define the WFs of two-noninteracting particles
confined in a harmonic well potential in one dimension, or
alternatively, they define WFs of a particle in a harmonic
potential in two-space dimensions. When $g\neq 0$, which indicates
that $\alpha $ would be greater than $1$, then the total WF
becomes
\begin{eqnarray}
&& \widetilde{W}_{l,n}\left( \overline{q},\overline{p};\overline{Q},\overline{P}%
\right) =\left( -1\right) ^{l}\ \exp \left[ -2\overline{Q}^{2}-2\overline{P%
}^{2}\right] \nonumber \\
&& \times \ L_{l}\left( 4\overline{Q}^{2}+4\overline{P}^{2}\right)
\frac{n!}{\Gamma \left( n+\alpha +1/2\right) }e^{-\overline{\omega }%
\overline{q}^{2}/2}\nonumber \\
&& \times L_{n}^{\alpha -1/2}\left[ \frac{\overline{\omega }}{2}%
\left( \overline{q}-i\ \partial \overline{_{p}}\right) ^{2}\right]
L_{n}^{\alpha -1/2}\left[ \frac{\overline{\omega }}{2}\left(
\overline{q}+i\
\partial _{\overline{p}}\right) ^{2}\right]\nonumber \\
&&\times \left[ \frac{\overline{\omega }}{%
2}\left( \overline{q}^{2}+\partial _{\overline{p}}^{2}\right)
\right] ^{\alpha }\ e^{-\overline{p}^{2}/2\overline{\omega }}.
\label{25}
\end{eqnarray}

Finally, by these considerations, we comment on
${\widetilde{W}}_n(\overline{q},\overline{p})$ given by Eq.
(\ref{24}), rather than Eq. (\ref{25}), to better visualize the
phase space behaviors of WFs presented in Figs. 1-2., i.e., how
the localization in $\overline{q}$ happens when the strength of
spatial confinement $\overline{\omega}$ is increased. The use of
the asymptotic expansion of the Laguerre polynomials for large
order \cite{Abro} yields Eq. (\ref{24}) to take form
\begin{eqnarray}\label{26}
\widetilde{W}_n(\overline{q},\overline{p})&\simeq &
\frac{(-1)^n}{\sqrt{\pi}}\left[ (n+\frac{1}{2})
(\overline{\omega}\,\overline{q}^2+\frac{\overline{p}^2}
{\overline{\omega}})\right]^{-1/4}\nonumber \\
&\times & cos\left\{ 2\left[
(n+\frac{1}{2})(\overline{\omega}\,\overline{q}^2+\frac{\overline{p}^2}
{\overline{\omega}})\right]^{1/2}-\frac{\pi}{4}\right\},
\end{eqnarray}
from which one can easily find out where the WF is zero and where
it takes negative values. For instance, Eq. (\ref{26}) has zeros
when
\begin{eqnarray}\label{27}
2\,\left[
(n+\frac{1}{2})(\overline{\omega}\,\overline{q}^2+\frac{\overline{p}^2}
{\overline{\omega}})\right]^{1/2}-\frac{\pi}{4} =
(k-\frac{1}{2})\pi,\qquad k=0,\pm 1, \pm 2,\dots ,
\end{eqnarray}
which clarifies the above comment that the localization in
$\overline{q}$ appears as $\overline{\omega}$ increases. Namely,
the left hand side of Eq. (\ref{27}) can be rearranged, in
dimensional units as usual, to give
\begin{eqnarray}\label{28}
\frac{2{\cal
H}(q,p)}{\omega}=\frac{\pi^2}{4(n+\frac{1}{2})}(k-\frac{1}{4})^2\hbar.
\end{eqnarray}
In fact, this last expression is called the symplectic area
enclosed by an ellipse whose boundary is given by ${\cal
H}(q,p)=E=p^2/(2m)+\omega^2 q^2/2$ and its minimum value is
determined by the Gromov's non-squeezing theorem, i.e., $2{\cal
H}(q,p)/\omega \geq \hbar$ \cite{Gosson}. Therefore, the
projections of WFs onto $(q,p)$ plane are elliptic energy shells
whose eccentricity is given by $e=\sqrt{1-(m\omega)^2}$, and they
are circles with the frequency $\omega=1/m$.

In this paper, we introduced and solved the WFs of one-dimensional
two particle Calogero-Sutherland system in which the particles
obeying the Boltzman statistics interact mutually by the sum of
quadratic and inversely quadratic pair potentials, and they are
confined in an external harmonic potential as well. It is obvious
that the technique introduced here can easily be extended to find
explicit analytical expressions for WFs of 3-and N-body
counterpart of the problem. Namely, by using Jacobi coordinates,
after separating the center-of-mass coordinate, one can easily
construct the remaining part of WF with $N-1$ relative
coordinates. Furthermore, due to the fact that, with particular
choices of the coupling constant $g$, the radical equation of
three dimensional isotropic oscillator and of hydrogen-like atom
in both spherical and parabolic coordinates, one dimensional three
body problem and the s-state of Morse potential\cite{Li05} are all
reduced to Calogero-Sutherland system, the results obtained here
unify inherently the WFs of these quantum mechanical problems.

As a final remark, we should point out that the attractive
interaction, i.e., $0>g\geq-\hbar^{2}/8\mu$, is also present in
the CSM. Therefore, one can easily compare the phase space
behaviors of WFs of two different regimes as well. In particular,
it should be noted  that a particular choice of $g$, when
$0>g\geq-\hbar^{2}/8\mu$, yields one dimensional band problem
solved by Scarf\cite{Sutherland71,Scarf58}. This serves as a model
for one dimensional dot arrays, as also indicated in the
Introduction section.

\begin{figure*}
\includegraphics*{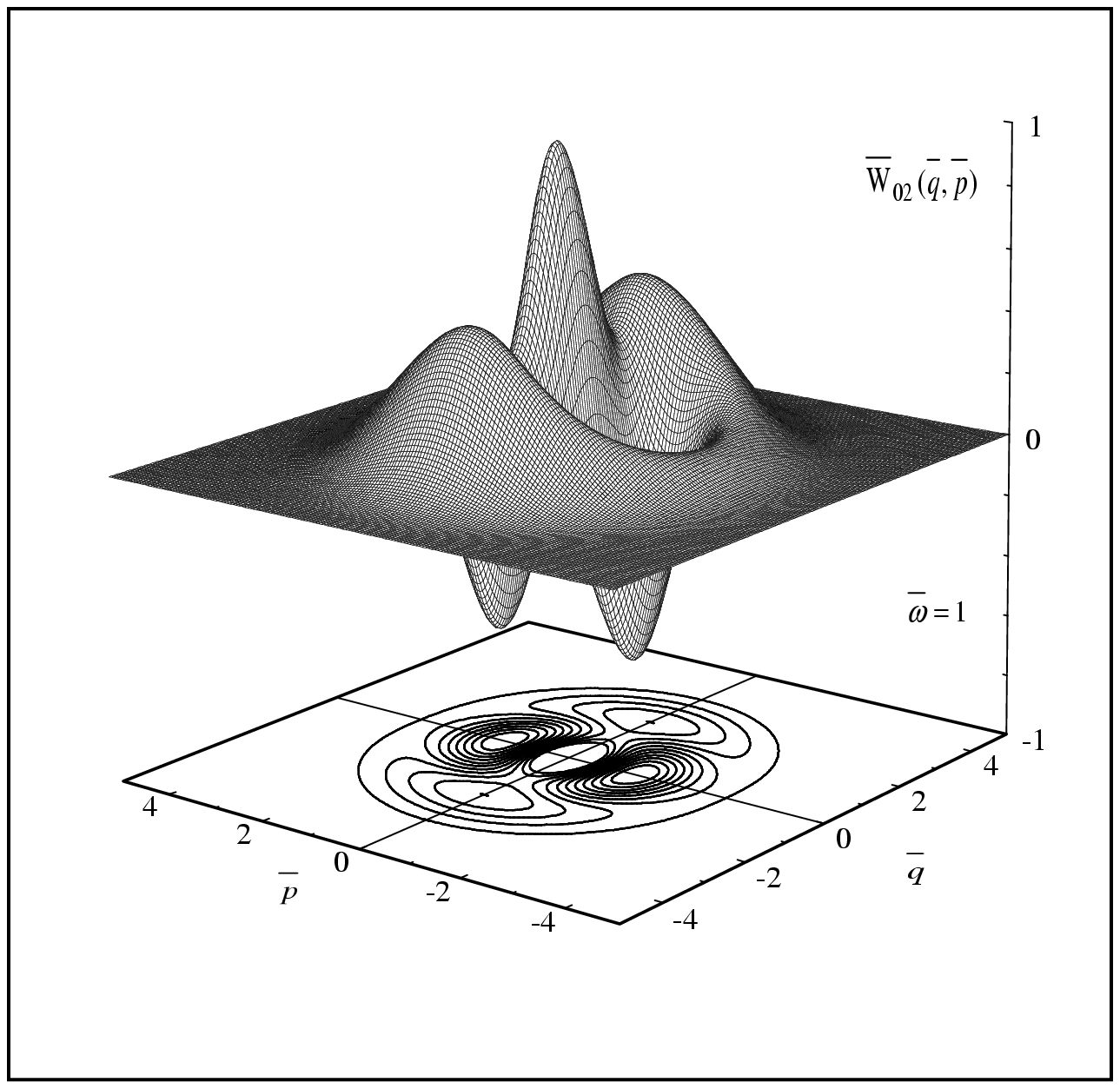} (a)
\end{figure*}
\begin{figure*}
\includegraphics*{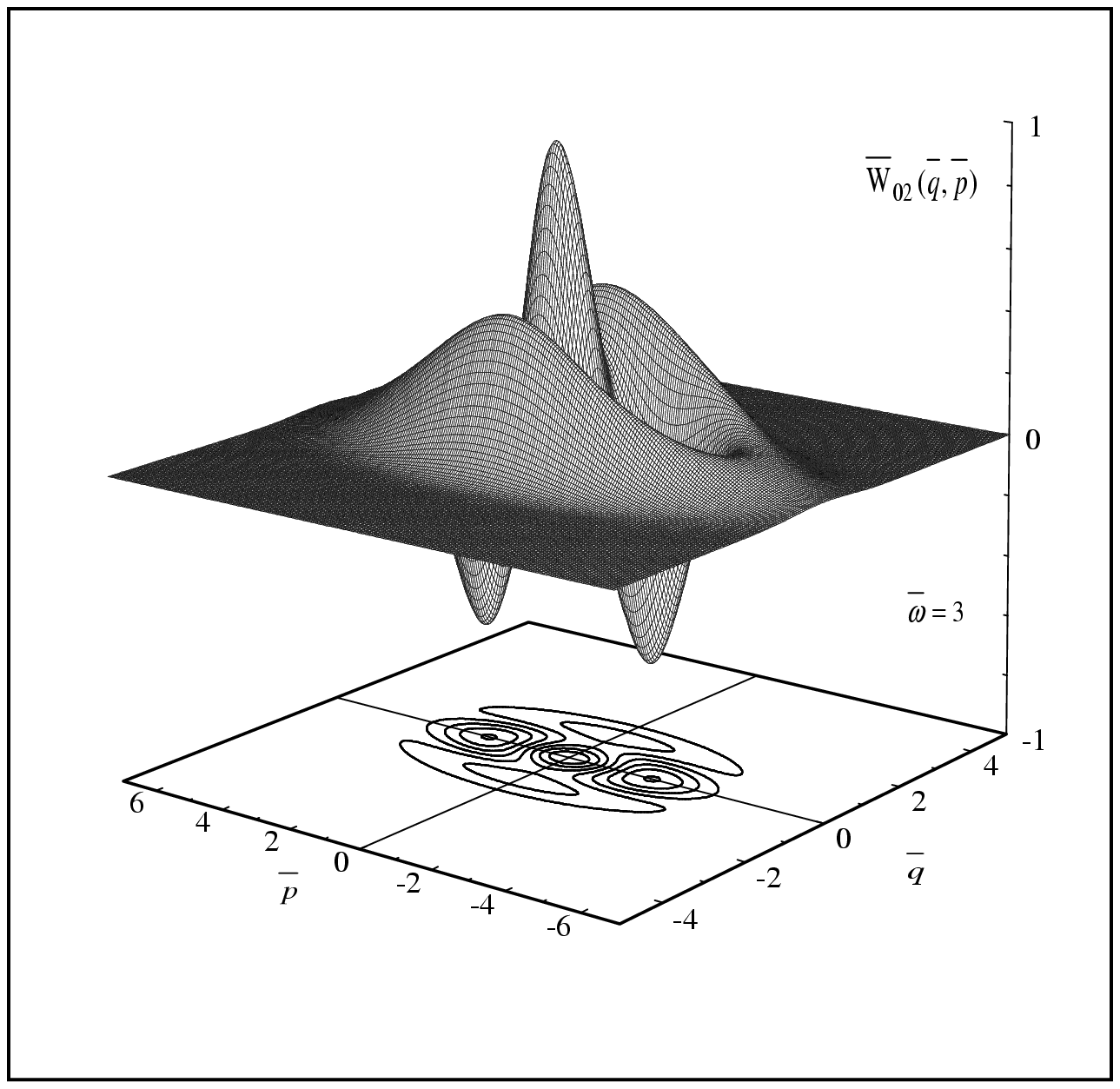} (b) %
\caption{Three dimensional plots of WF $\widetilde{W}_{02 }$ for
the relative motion (Eq.~(\protect\ref{19})) as a function of
dimensionless position $\overline{q}=q/l$ and momentum
$\overline{p}=lp/\hbar $ for
dimensionless frequency (a) $\overline{\protect\omega }=\protect\omega /%
\protect\omega _{\bullet }=1$ and (b) $\overline{\protect\omega
}=3$.}
\end{figure*}

\begin{figure*}
\includegraphics*{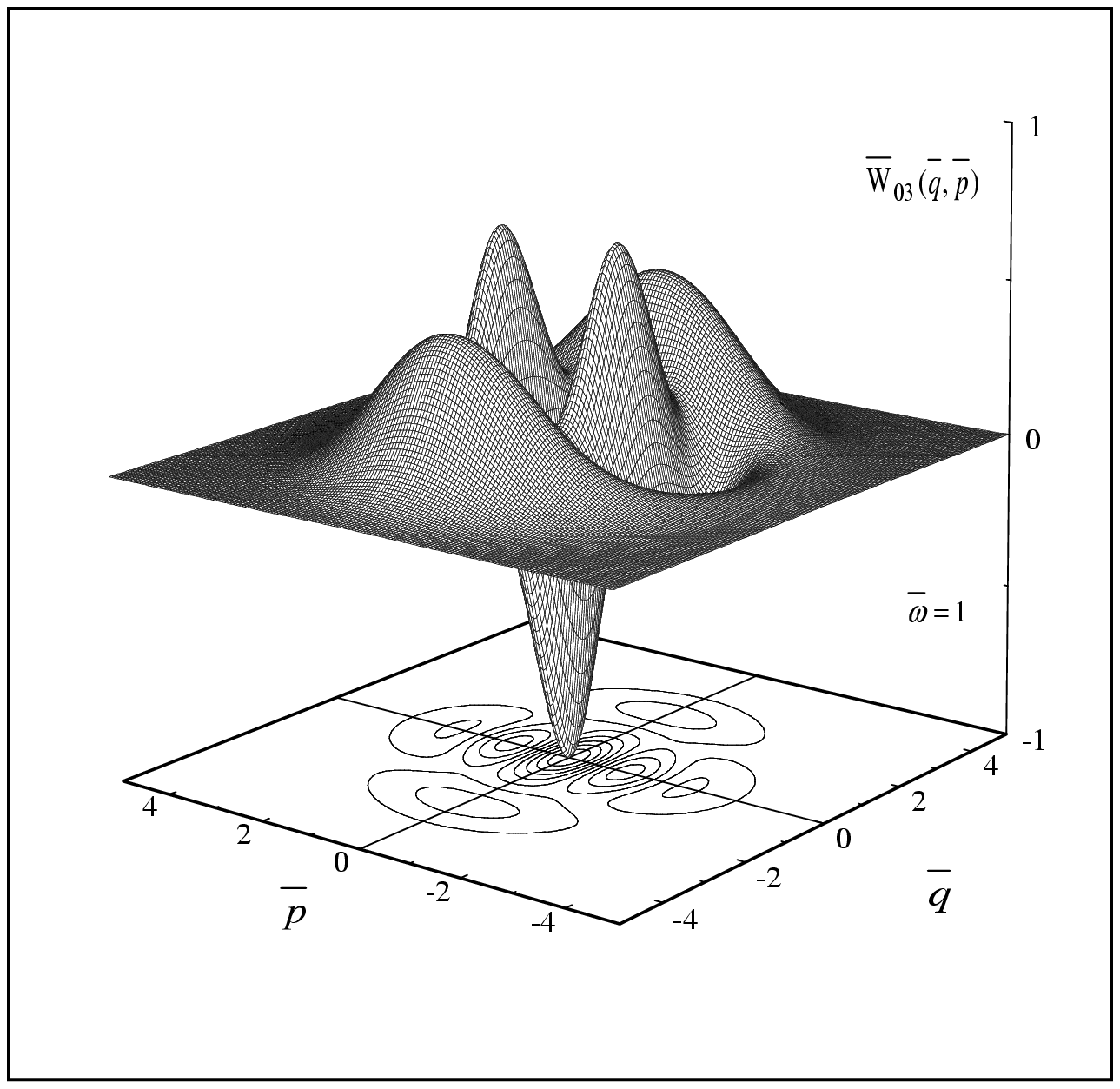} (a)
\end{figure*}
\begin{figure*}
\includegraphics*{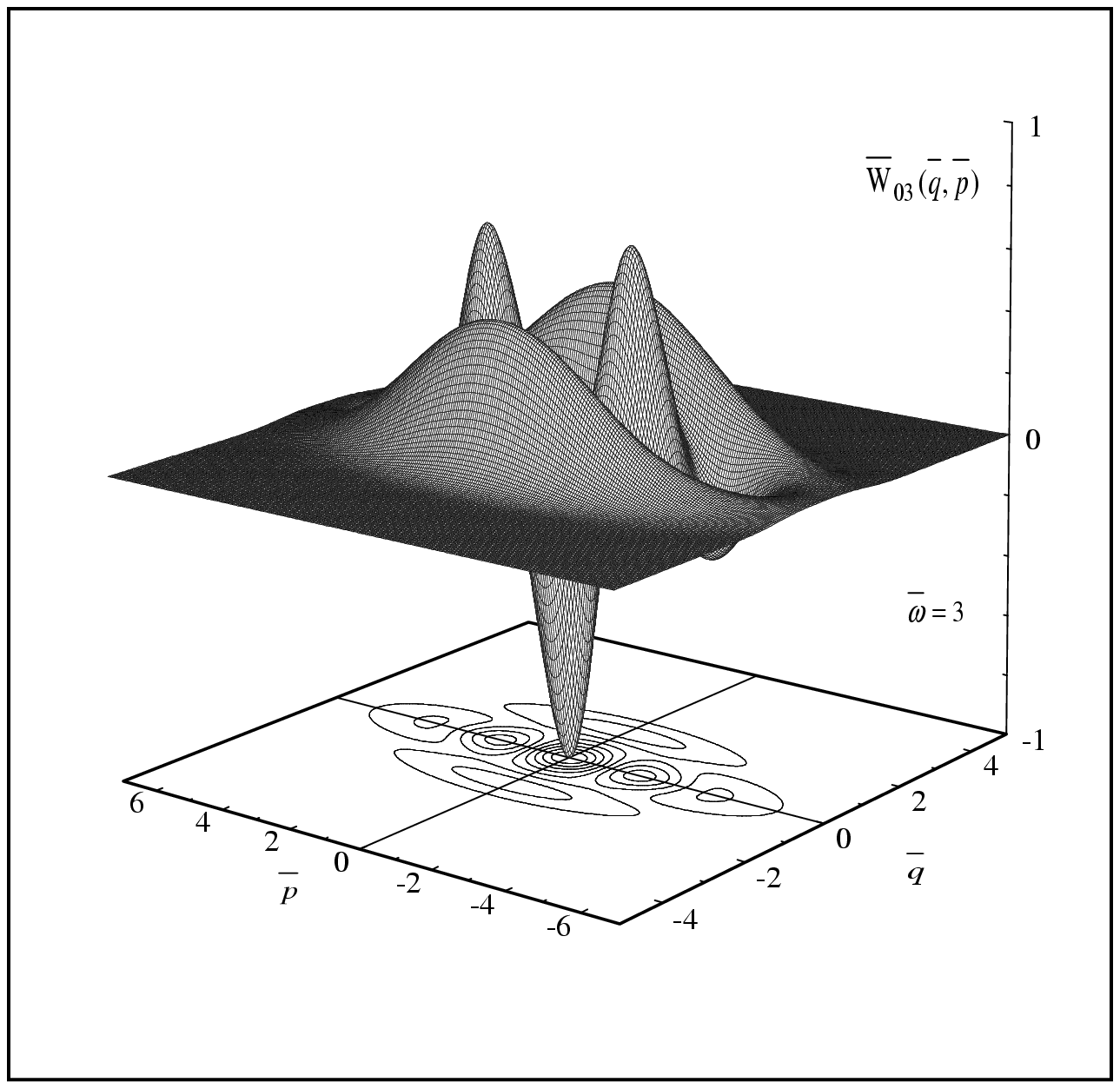} (b)
\caption{Three dimensional plots of WF $\widetilde{W}_{03 }$ for
the relative motion (Eq.~(\protect\ref{19})) as a function of
dimensionless position $\overline{q}=q/l$ and momentum
$\overline{p}=lp/\hbar $ for
dimensionless frequency (a) $\overline{\protect\omega }=\protect\omega /%
\protect\omega _{\bullet }=1$ and (b) $\overline{\protect\omega
}=3$.}
\end{figure*}

\end{document}